\documentclass[%
 reprint,
superscriptaddress,
 amsmath,amssymb,
 aps,
]{revtex4-1}

\usepackage{graphicx}
\usepackage{adjustbox}
\usepackage{dcolumn}
\usepackage{bm}
\usepackage{xcolor,colortbl}
\usepackage{float}
\setcitestyle{numbers,sort&compress}
\usepackage[colorlinks=true, allcolors=blue]{hyperref}
\usepackage[colorinlistoftodos,textsize=tiny]{todonotes}
\usepackage{mathtools} 
\usepackage{overpic}
\graphicspath{ {./images/} }

\begin{document}

\preprint{APS/123-QED}
\title{Enhanced diffusion of colloidal tracers due to enzymatic activity}

\author{Mauricio Gomez}
\thanks{equally contributing authors, listed alphabetically}
\affiliation{Department of Physics,  California State University Fullerton, CA 92831 USA}
\affiliation{Lewis-Sigler Institute for Integrative Genomics, Princeton University, Princeton, NJ, 08544, USA} 

\author{Erick Leyva}
\thanks{equally contributing authors, listed alphabetically}
\affiliation{Department of Physics,  California State University Fullerton, CA 92831 USA}

\author{Justine Miqueu-Petit}
\thanks{equally contributing authors, listed alphabetically}
\affiliation{Universit\'{e} de Toulouse, CNRS, Laboratoire de Physique Th\'{e}orique, Toulouse, France} 

\author{Dakota Feldcamp}
\affiliation{Valencia High School, Placentia, CA 92870 USA}

\author{Anthony Estrada}
\affiliation{Department of Physics,  Syracuse University, Syracuse, NY 13244 USA}

\author{W.~Benjamin Rogers}
\affiliation{Department of Physics,  Brandeis University, Waltham, MA 02452 USA}

\author{Jennifer L.~Ross}
\affiliation{Department of Physics,  Syracuse University, Syracuse, NY 13244 USA}

\author{Wylie W.~Ahmed}
\email[correspondence: ]{wylie.ahmed@utoulouse.fr }
\affiliation{Department of Physics,  California State University Fullerton, CA 92831 USA}
\affiliation{Universit\'{e} de Toulouse, CNRS, Laboratoire de Physique Th\'{e}orique, Toulouse, France} 
\affiliation{Universit\'{e} de Toulouse, CNRS, Centre de Biologie Int\'{e}grative, Toulouse, France}

\begin{abstract}
Enzymatic catalysis can generate nonequilibrium fluctuations, but how these couple to tracer motion at larger length scales depends on physical context. Here, we investigate colloidal tracers in two configurations: passive particles dispersed in an enzymatically active solution, and enzyme-decorated particles where catalysis occurs directly at the tracer surface. We combine differential dynamic microscopy (DDM), which probes ensemble-averaged long-time diffusion, with optical tweezer (OT) measurements of short-time force fluctuations, and compare several complementary metrics for quantifying activity-induced enhancement. For 1~$\mu$m tracers, we observe activity-induced enhancements in both configurations, with the strongest effects for enzyme-decorated particles, which exhibit enhanced diffusion and increased non-thermal force fluctuations. For 200 nm tracers, enhancements are more subtle and method-dependent: DDM detects modest increases in diffusion for bare particles, while corresponding signatures are not resolved by the OT. These results demonstrate that enzymatic activity can be transduced from molecular to microscale motion and forces, but that the apparent magnitude and detectability of enhancement depend strongly on tracer size, localization of activity, the timescales probed by the measurement, and the metric used to quantify enhancement. More broadly, understanding how enzyme activity modifies transport and fluctuations across scales is important for interpreting nonequilibrium dynamics in active soft matter, intracellular transport, and chemically crowded biological environments.

\end{abstract}

\maketitle

\maketitle

\section{Introduction}

Enzymatic reactions have been proposed as a source of non-thermal forces capable of altering the motion of enzymes themselves or nearby tracer particles~\cite{ghosh2021enzymes, jee2020master, jee2018catalytic, patino2018fundamental}. Early reports suggested large increases in enzyme diffusivity during catalytic turnover~\cite{jee2020master, Xu2019, Muddana2010, Riedel2015, Sengupta2013}, raising the possibility that chemical activity could be harnessed to drive fluctuations in soft matter systems. However, subsequent studies questioned both the magnitude and the existence of such effects~\cite{Chen2020, Gunther2019, choi2022displacement}, and enhanced enzyme diffusion is still actively debated~\cite{zhang2019enhanced, feng2020enhanced, presse2020thermodynamic}.  

For example, single-molecule tracking experiments by Choi and Xu~\cite{choi2022displacement} found no measurable enhancement of freely diffusing enzymes at short timescales (600 $\mu$s). In contrast, other single-molecule tracking studies~\cite{Xu2019, scott2026enhanced} conducted in constrained 2D motion in lipid bilayers observed enhanced diffusion at longer timescales ($\sim 1$ s). A recent study of tethered single molecule nanomotors shows that enzymes can indeed transduce chemical to mechanical energy~\cite{tang2025single}. Theoretical analyses have further shown that viscosity and hydrodynamic coupling place limits on the magnitude of possible enhancements~\cite{golestanian2015enhanced, adeleke2019chemical, alston2026stochastic, tripathi2022gauging}, hypothesize the microscopic mechanisms~\cite{agudo2018phoresis, astumian2014enhanced, illien2017diffusion}, and their macroscopic~\cite{ichii2026enzyme} consequences.

Beyond single enzymes, enzyme-decorated colloids have been investigated as a means to amplify catalytic activity. Dey \emph{et al.}~\cite{dey2015micromotors}, for example, reported a 22\% increase in the diffusion coefficient of urease-coated particles measured by single particle tracking. Optical tweezer measurements have provided direct evidence that enzymatic reactions can generate measurable forces in enzyme-decorated colloids. Ma \emph{et al.} measured an effective force of $\sim64$~fN for $\sim390$~nm catalase-powered hollow mesoporous silica Janus nanomotors, using a force-spectrum analysis~\cite{ma2015enzyme}. Patiño \emph{et al.} later measured a larger force of $\sim170$~fN for 2~$\mu$m urease-powered PS@SiO$_2$ micromotors and showed that both speed and force depend nonlinearly on enzyme coverage~\cite{patino2018influence}. Together, these studies show that enzyme identity, colloid architecture, enzyme loading, and particle size all influence how molecular-scale catalytic activity is transduced into colloidal motion and force. Recent work in nanoscale DNA origami further supports this view, showing that propulsion depends on enzyme number, spatial arrangement, catalytic loading, and geometric anisotropy in a non-monotonic manner~\cite{paffen2025programmable}.

In this work, we examine how molecular-scale enzymatic activity influences the motion of larger tracer particles, and under what conditions such effects become detectable. While prior studies have suggested that enzymatic catalysis can generate nonequilibrium fluctuations, the extent to which these fluctuations are transduced to micron-scale tracers remains context-dependent. We compare two particle sizes (200 nm and 1 $\mu$m) and two tracer configurations. In the \emph{bare case}, tracers are passive colloids dispersed in an enzymatically “active bath”, where any enhancement must arise from collisions or hydrodynamic coupling to enzymes free in solution. In the \emph{decorated case}, urease molecules are directly tethered to the tracer surface, so that chemical activity occurs locally at the particle interface, effectively turning the tracer into an “active particle”. These two configurations allow us to distinguish between activity that originates from the bulk solution and activity that is localized to the tracer itself.

A central goal of this work is not only to assess the presence of activity-induced motion, but also to demonstrate how different measurement modalities and definitions of enhancement influence its detection and interpretation. To probe tracer dynamics across complementary timescales, we combine differential dynamic microscopy (DDM)~\cite{cerbino2008differential}, which measures ensemble-averaged displacements over longer times ($\mathcal{O}(10^{-2} - 10^1~\mathrm{s})$), with direct force measurements using optical tweezers (OT)~\cite{farre2010force}, which probe single-particle force fluctuations at shorter times ($\mathcal{O}(10^{-4} - 10^0~\mathrm{s})$). Importantly, activity-induced enhancement can be quantified in multiple ways, each emphasizing different aspects of the underlying nonequilibrium dynamics. Before presenting the experimental results, Sec.~III introduces the enhancement metrics used throughout this work and discusses their underlying assumptions.


\begin{figure}[th]
    \includegraphics[width = 0.47 \textwidth]{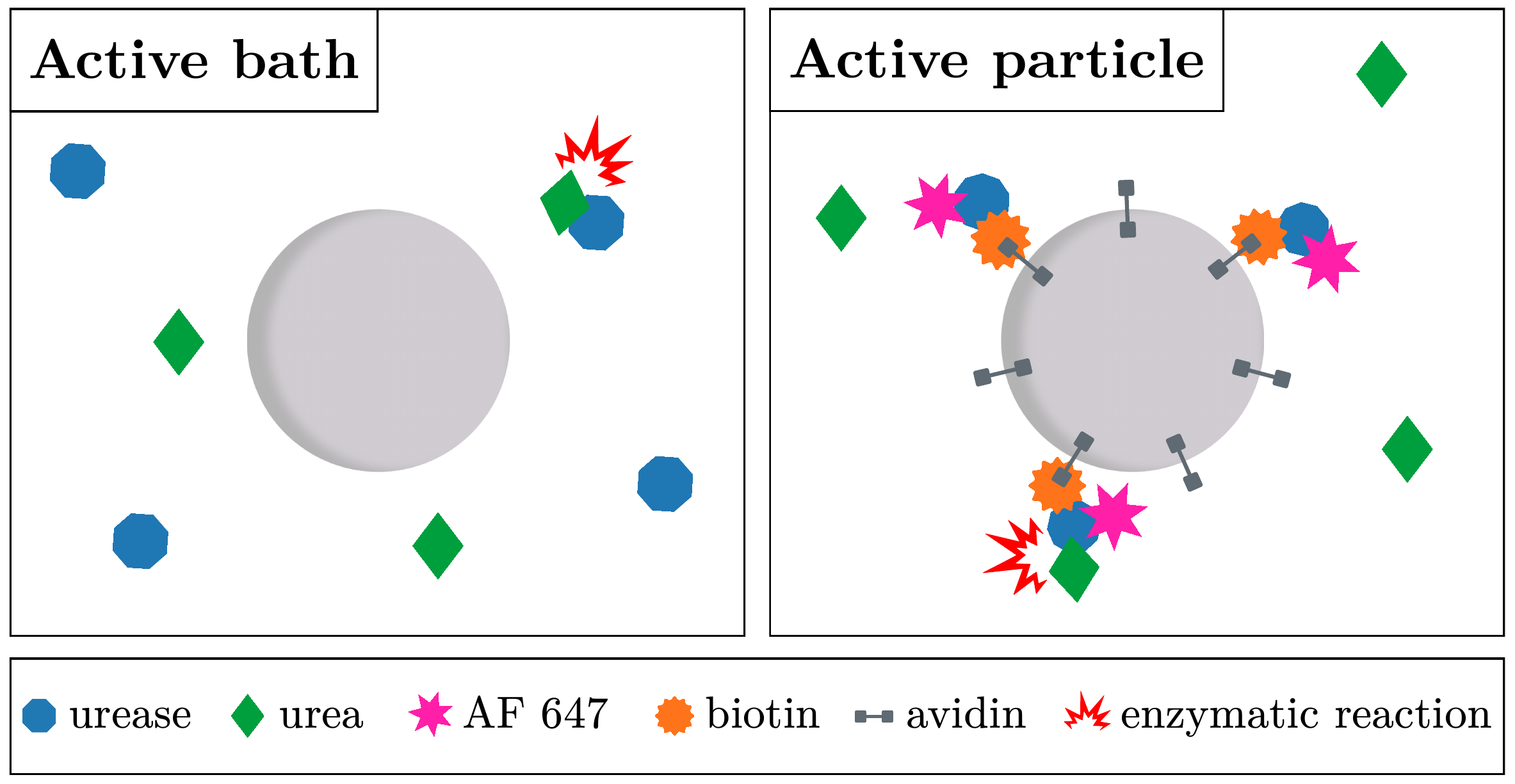}
    \caption{(Left) In an active bath, passive colloidal tracers (gray) are suspended in a viscous buffer containing freely diffusing urea and urease, where enzymatic activity occurs in the bulk solution.  These are referred to as ``bare particles''. (Right) In the active particle configuration, urease is fluorescently labeled and tethered directly to the colloid via biotin–avidin binding, localizing catalytic activity at the particle surface. The surrounding buffer contains urea to sustain the enzymatic reaction at the colloid surface.  These are referred to as ``decorated particles''. Schematic not drawn to scale.}
    \label{fig:EnzymeSetup}
\end{figure}

\section{Materials and Methods}

\subsection{Materials}

Urease from Jack Bean was purchased from TCI Chemicals (U0017). For selected experiments, urease was fluorescently labeled with Alexa647 and biotinylated using commercial kits (Molecular Probes Fluorescent Protein Labeling and EZ-Link Sulfo-NHS-LC-Biotinylation, Thermofisher, 21435). Urea was obtained from Sigma Aldrich (U5375). Tracer particles were neutravidin-coated polystyrene beads of diameter 200 nm and 1 micron (Invitrogen, F8811 and F8823). Buffer solutions included 1 M HEPES (Sigma Aldrich, BP299-100) and 10× DPBS (Gibco, 14200-075). Phenol red (Sigma Aldrich, P3532) was used to monitor enzymatic activity. 

\subsection{Sample preparation of enzyme baths}

Passive baths were prepared by mixing 1 mL of buffer (10× DPBS for DDM experiments and 1M HEPES for trapping experiments) with 300 mM urea and tracer beads. Active baths were prepared identically to passive baths with the addition of urease at a concentration of 50 U/mL. In this configuration, tracer particles remained chemically inert (“bare”), and any enhancement of tracer motion originated from enzymatic reactions in the bulk solution. Passive and active baths contained either 200 nm or 1 $\mu$m beads at dilutions of 1:100–1:50,000 from original stock solution.  High concentrations of tracer particles were used for DDM and lower concentrations for OT.

To ensure the activity of our urease, an activity assay was utilized following established protocols~\cite{okyay2013high}. Briefly, the protocol uses ammonia production in the urease-urea reaction to detect a shift in the absorption spectra with a pH indicator. We used UV-Vis spectroscopy to measure the absorbance shift caused by a pH indicator. We mixed 300 mM urea and 100$\mu$M phenol red with either 10X DPBS or 1M HEPES. We then added urease at various concentrations. We found that for a urease concentration of  50 U/mL, the reaction continued for 10-20 min, after which the absorbance reached a steady plateau value, consistent with established pH dependence of urease activity~\cite{panja2021urea}. All experiments were conducted during the time window where the urease reaction was active.

\subsection{Urease-decorated (active) particles}

\begin{figure}[th]
    \includegraphics[width = 0.45 \textwidth]{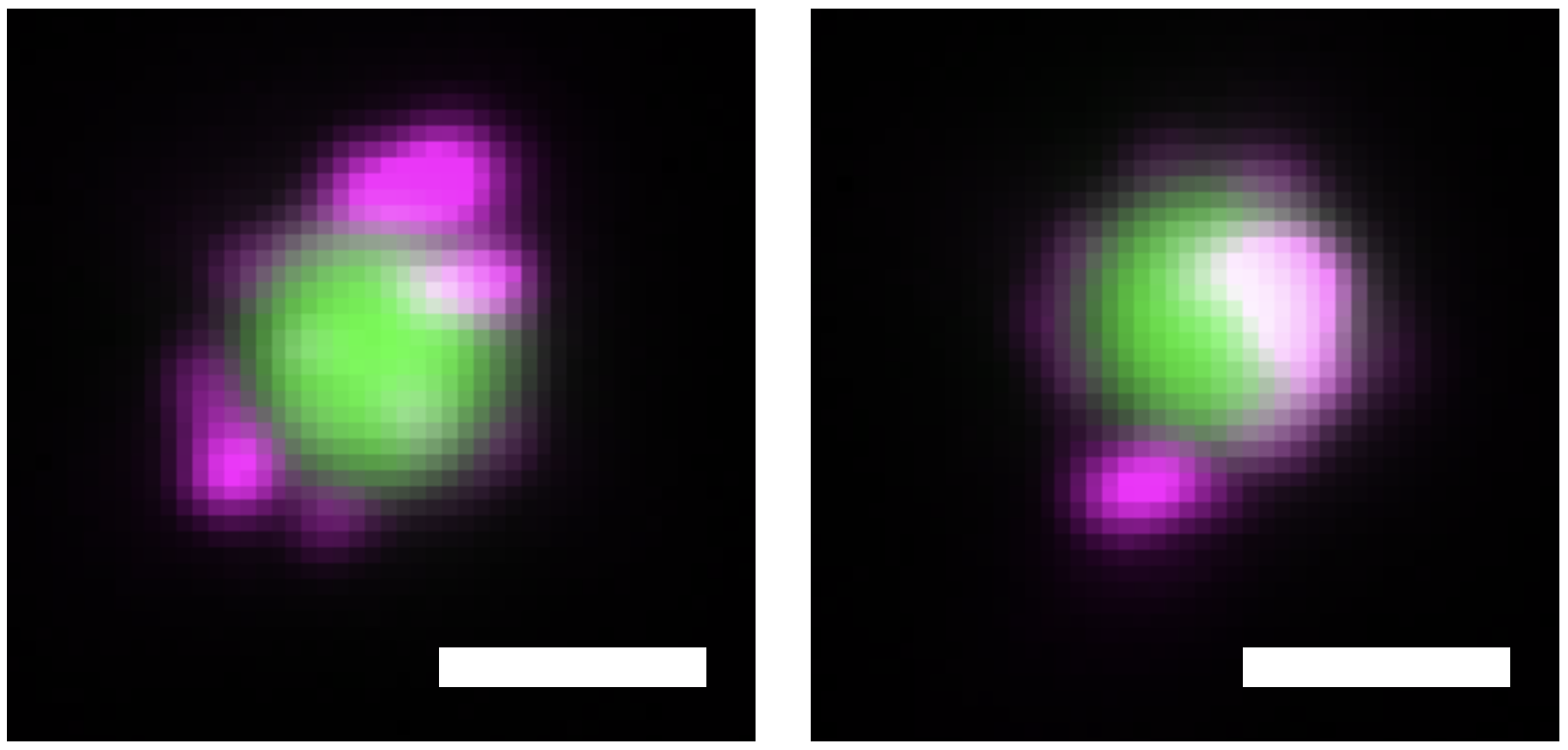}
    \caption{Representative images of urease-decorated active particles where urease is magenta and the tracer colloid is green.  Scale bar = 1 $\mu$m}
    \label{fig:ActiveParticle}
\end{figure}

To create enzyme-decorated tracers, urease was biotinylated and fluorescently labeled using NHS-ester chemistry. Neutravidin-coated beads were incubated with the biotinylated urease at a 1:10 bead-to-protein ratio for 2 hours at room temperature. Following incubation, the beads were washed twice to remove unbound enzyme. The resulting urease-decorated particles were resuspended in 1 M HEPES buffer. Confocal fluorescence imaging confirmed urease attachment on the bead surfaces (see Fig.~\ref{fig:ActiveParticle}).  


In the absence of the urea substrate, these suspensions were considered ``passive decorated'' samples. Addition of 300 mM urea created “active decorated” samples, in which catalytic activity was localized at the tracer surface. These active particles therefore represent a distinct configuration from bare tracers in active baths, as enzymatic forces are coupled to the colloid itself rather than originating from the bulk solution.  

\subsection{Preparation of sample chambers}

Sample chambers were assembled from a standard glass microscope slide and a No.~1.5 coverslip. The sample thickness was approximately 200 $\mu$m, created by either spacers or vaccum grease. This simple static chamber design was used for both DDM and OT measurements on the same microscope.

\subsection{Microscopy}

Microscopy and optical trapping experiments were performed simultaneously on a Nikon TE2000 inverted microscope equipped with a 60×/1.2 NA water-immersion objective and a Hamamatsu ORCA-Flash4.0 V2 sCMOS camera. All videos were recorded at 100 frames per second with a field of view of 1024 × 1024 pixels for a duration of 20-30 s. Each trial was an image sequence of 2000-3000 frames.


\subsection{Optical Trapping}
The optical tweezer system (Impetux Optics S.L.) is built on the same microscopy system as above and includes the optical trap, piezo stage positioning, and force detection. The 60x/1.2NA objective focuses the near-infrared laser (1064 nm, IPG-YLR-10, IPG Photonics) to create the optical trap. The photon momentum method (PMM)~\cite{farre2010force, gieseler2020optical} was implemented with a 1.4NA oil immersion condenser and a position sensitive sensor, digitized at 50 kHz, and allows for force detection and laser tracking interferometry. For the force calibration to be accurate, it is critical to use a condensing objective with higher numerical aperture than the trapping objective and to minimize scattering of light through the sample~\cite{farre2010force,jun2014calibration}. Labview (National Instruments) was used to control all experimental hardware and data acquisition. Trap stiffness for 1 $\mu$m bare and decorated particles was 83.6 and 81.1 pN/$\mu$m, respectively. For 200 nm particles, the trap stiffness was 4.8 pN/$\mu$m for bare particles and 6 pN/$\mu$m for the decorated particles.

\subsection{DDM analysis}

Tracer diffusivity was quantified using differential dynamic microscopy (DDM) following standard procedures~\cite{cerbino2017perspective, verwei2022quantifying}. Image sequences were analyzed to obtain the intermediate scattering function, from which effective diffusion coefficients were extracted by fitting the relaxation times as a function of wavevector. Results were averaged over multiple fields of view, and control measurements with passive colloids  were used to verify analysis parameters.

\subsection{OT Force Data Analysis}

Force signals $\mathbf{F}(t)$ were sampled at 50~kHz, and their power spectra $\langle |\tilde{\mathbf{F}}|^2 \rangle$ were computed using Welch’s method with a Hamming window~\cite{welch1967use}. Each spectrum was fit with a Lorentzian~\cite{berg2004power} by nonlinear least squares~\cite{coleman1996interior} to estimate the effective particle radius. When a Lorentzian fit estimated a particle radius outside one standard deviation of the distribution peak, the measurement was flagged as likely corresponding to multiple trapped particles or aggregates --- this issue occured primarily with 200 nm particles. Obvious aggregates were excluded from the OT analysis. However, small unresolved clusters or transient aggregation may still affect the effective hydrodynamic radius. The active force spectra, quantifying only non-thermal fluctuations, was obtained by $\langle |\tilde{\mathbf{F}}_{\mathrm{nonthermal}}|^2 \rangle = \langle |\tilde{\mathbf{F}}_{\mathrm{active}}|^2 \rangle - \langle |\tilde{\mathbf{F}}_{\mathrm{passive}}|^2 \rangle$.

\subsection{Viscosity of buffer solutions}
To determine the viscosity of buffer solutions, a 6 $\mu$m polystyrene bead was dragged at a constant speed of 60 $\mu$m/s. By measuring the force on the trapped particle directly, we then used Stokes law to calculate the viscosity~\cite{pesce2005viscosity}. We found that the solutions made with 10X DPBS and 1M HEPES had viscosities of 1.4 mPa-s and 2.0 mPa-s, respectively. As a control, we measured the viscosity of water which was found to be 0.98 mPa-s.

\subsection{Theoretical model}
Many nano- and microscopic mechanisms have been proposed to explain enhanced enzyme diffusion~\cite{tripathi2022gauging, feng2020enhanced, agudo2018phoresis, illien2017diffusion, golestanian2015enhanced}. Here, our goal is not to identify the microscopic origin of the enzymatic forcing, but rather to describe how a generic nonthermal active fluctuation would appear in the force spectrum of a trapped colloidal tracer. We therefore adopt a minimal mesoscopic Langevin description of colloidal dynamics~\cite{fodor2018statistical,sekimoto1998langevin}, modeling the nonthermal component as an Active Ornstein-Uhlenbeck process, as in our previous work~\cite{jones2021stochastic}.

The position of the trapped particle is governed by 

\begin{equation}
    \gamma \dot{\mathbf{r}} + \kappa \mathbf{r} = \gamma \mathbf{u} + \sqrt{2D} \gamma \boldsymbol{\xi}
    \label{eq:eom}
\end{equation}
which balances the deterministic frictional and optical trap forces with the random active and thermal forces. In equation (\ref{eq:eom}), $\mathbf{r}$ is the colloid position, $\mathbf{\dot{r}}$ is the velocity,  $\kappa$ is the optical trap stiffness, $\gamma$ is the friction coefficient corresponding to Stokes' drag, $D$ is the thermal diffusion coefficient, and $\mathbf{\xi}$ is the zero mean $\delta-\text{correlated}$ Gaussian white noise process. The nonthermal activity is modeled with an effective active burst velocity, $\mathbf{u}$. The burst velocity is modeled as an Active Ornstein-Uhlenbeck (AOUP) process with characteristic strength $\vert \mathbf{u} \vert = u$, and timescale $\tau$ ~\cite{fodor2018statistical,martin2021statistical}.  This model yields an analytical form of the force spectrum, previously derived in~\cite{jones2021stochastic},

\begin{equation}
S\!_{f\!f}(\omega) = \frac{2 \kappa^2 D}{\mu^2 + \omega^2} + \frac{2 \kappa^2 \tau u^2}{(\mu^2 + \omega^2)(1 + \omega^2 \tau^2)}
\label{eq:forcespectrum}
\end{equation}
where the first and second term represent the thermal force spectrum and the nonthermal force spectrum, respectively.

\section{Quantifying enhanced diffusion}

Enhanced diffusion can be quantified in multiple ways, even from the same underlying data. In DDM, enhancement is often quantified through an effective diffusion coefficient, but deviations from purely diffusive behavior can also be captured, for example, through ballistic-like contributions where an effective velocity is defined. In OT measurements, enhancement can be quantified in several complementary ways, including model-dependent estimates of long-time diffusion, fitting the nonthermal force spectrum to extract parameters, direct comparison of force spectra, integration of the nonthermal spectral power, and analysis of the variance of force fluctuation distributions.

These different metrics are not redundant, but instead probe distinct aspects of nonequilibrium fluctuations and are sensitive to different timescales. In this section, we outline the measures used in our analysis.

\subsection{Difference in diffusion, $\Delta D/D_p$}

A common metric for quantifying activity-induced enhancement is the relative change in diffusion, $\Delta D / D_p$, where $\Delta D = D_a - D_p$ is the difference between the effective (active) and thermal (passive) diffusion coefficients. This measure has been widely used in the enhanced enzyme diffusion literature~\cite{zhang2019enhanced, Xu2019}.

In DDM, each image sequence contains hundreds to thousands of particles, which are analyzed in Fourier space to extract a single ensemble-averaged diffusion coefficient per dataset. Repeating this analysis across multiple independent measurements yields distributions of diffusion coefficients for both the active ($D_a$) and passive ($D_p$) conditions.

To obtain a comparable estimate from OT measurements, we first fit the passive force spectrum to the thermal part of the analytic model in equation (\ref{eq:forcespectrum}) to extract the trap stiffness $\kappa$ and friction coefficient $\gamma$, which defines the thermal diffusion coefficient $D_p$. These parameters are assumed to be the same in passive and active experiments and thus remain fixed. Then the force spectrum of the active experiments is fit with an effective (scalar) temperature as the only free parameter. This effective temperature is used to compute the corresponding active diffusion coefficient $D_a$.

This approach assumes Brownian-like dynamics where fluctuations (both thermal and nonthermal) are frequency-independent.

\subsection{Estimating force from long-time diffusion, $F$}

The effective driving force associated with activity-induced fluctuations can be estimated from the long-time translational diffusion coefficient~\cite{ma2015enzyme}. For a spherical particle with two orientational degrees of freedom in a harmonic potential~\cite{ten2011brownian}, the force is given by
\begin{equation}
    F = \frac{3k_B T}{2R}\sqrt{2\left(\frac{D_a}{D_p} - 1\right)}.
\end{equation}
where $D_a \geq D_p$. This approach also assumes Brownian-like dynamics, and can be computed for both DDM and OT data once diffusion coefficients are calculated.

\subsection{Fitting of nonthermal force spectrum, $u^2 \tau$}

In OT measurements, we directly access the nonthermal force spectrum, defined as the difference between the force spectra in the active and passive conditions. Within our model, the trap stiffness and friction coefficient are assumed to be unchanged between these conditions. We then fit the active component of the analytical force spectrum to the measured nonthermal spectrum in equation (\ref{eq:forcespectrum}) to extract the characteristic strength $u$ and timescale $\tau$ of the active process. From these parameters, the corresponding active diffusion coefficient is obtained as $D_a = u^2 \tau$. This approach assumes there is a well-defined amplitude and dominant timescale of the active process~\cite{fodor2018statistical, martin2021statistical}.


\subsection{Variance of force fluctuations distribution, $\Sigma$}

In OT measurements, we directly measure force fluctuations and can estimate their probability distributions, as done previously~\cite{patino2018influence, jones2021stochastic}. For both active and passive conditions, we compute the distributions and their mean and variance ($\mu$, $\sigma^2$). We then quantify enhancement by taking the ratio of variances, $ \Sigma = {\sigma_a^2} / {\sigma_p^2} -1$. This metric does not rely on a specific model, but instead provides a simple Gaussian-based measure for comparing the magnitude of fluctuations between conditions.  This approach assumes the active process is symmetric and stationary and does not consider higher order moments. Related fluctuation-based approaches have been used for urease-powered micromotors, where Patiño \emph{et al.} estimated propulsion forces from changes in the distribution of trapped-particle fluctuations~\cite{patino2018influence}.

\subsection{Ratio of force spectrum, $E_a$ and $J$}

The ratio of active to passive force spectra can be used to define an effective, frequency-dependent energy scale~\cite{seyforth2022nonequilibrium, gallet2009power, ben2011effective}. By subtracting the thermal baseline ($k_B T$), this yields a nonthermal energy spectrum, $E_a(\omega)$, which quantifies activity-induced fluctuations as a function of frequency. Integrating this quantity over the full accessible frequency range (1-25 kHz) (considering only positive contributions) defines an energy dissipation rate, $J = \int E_a(\omega)\, \mathrm{d}\omega$ in units of $k_B T / \mathrm{s}$.  
This approach assumes a Newtonian solvent, although extensions to systems with memory effects are possible~\cite{fodor2016nonequilibrium}.

\subsection{Integration of nonthermal force spectrum, $\mathbf{S}$}

The nonthermal force spectrum, $\vert \mathbf{\tilde{F}}_{\mathrm{nonthermal}} \vert ^2$, defined as the difference between the active and passive force spectra, provides a model-independent measure of the additional forces injected into the system beyond thermal fluctuations (within the measured translational degrees of freedom). Integrating this spectrum over the frequency range of the measurements (1-25 kHz),
$
    \mathbf{S}=\int \vert \mathbf{\tilde{F}}_{\mathrm{nonthermal}} \vert ^2 \, \mathrm{d}\omega,
$
and taking the square root yields a scalar estimate of the magnitude of the activity-induced driving force. Because this quantity is computed directly from experimental spectra without fitting, it provides a model-free estimate of the additional forces arising from enzymatic activity. This approach is similar to the PSD-based OT analysis of Ma \emph{et al.}, where the difference between force spectra with and without fuel was integrated to estimate the effective driving force of a single enzyme-powered nanomotor~\cite{ma2015enzyme}.

\section{Results}

To assess whether enzymatic activity influences tracer motion, we systematically examined the diffusion and force fluctuations of colloidal particles of two different sizes (200 nm and 1 $\mu$m) in both bare (active bath) and urease-decorated (active particle) configurations. For each case, we present results obtained from differential dynamic microscopy (DDM), which captures ensemble-averaged diffusion dynamics, and optical tweezer (OT) measurements, which directly probe single-particle force fluctuations across shorter timescales. 

We first present results for 1 $\mu$m tracers, where activity-induced effects are most pronounced, followed by results for 200 nm tracers, where such effects are weaker or absent. Within each size, we compare bare and enzyme-decorated configurations to isolate the role of activity localization.  All resulting enhancements are tabulated in Tables \ref{tab:DDM_enhancement_table} and \ref{tab:OT_enhancement_table}.

\begin{table}[h!]
\centering 
\setlength{\tabcolsep}{10pt}
\renewcommand{\arraystretch}{1.3}
\begin{adjustbox}{max width= 0.25
\textwidth}
\begin{tabular}{lcccc}
\hline
  & $\Delta D /D_p ($\%$)$
  & $F(\mathrm{fN})$ \\
\hline
1 $\mu m$ bare & 9.6 & 5.4 \\
1 $\mu m$ dec  & 14.3 & 6.6   \\
$200$ nm bare & 10 & 27.6 \\ 
$200$ nm dec & -7.9 & -- \\

\hline
\end{tabular}
\end{adjustbox}
    \caption{\textbf{Enhancement via Differential Dynamic Microscopy measurements.} Tabulated enhancement metrics for all conditions. A negative value indicates that the fitted active diffusivity was lower than the passive value; for the 200 nm decorated particles, this comparison is likely dominated by aggregation rather than an activity-induced reduction in diffusion. Entries marked (--) indicate a value that is not accessible.
     }
    \label{tab:DDM_enhancement_table}
\end{table}

\begin{table}[h!]
\centering 
\setlength{\tabcolsep}{8pt}
\renewcommand{\arraystretch}{1.2}
\begin{adjustbox}{max width=0.45\textwidth}
\begin{tabular}{lcccccc}
\hline
 & $\Delta D/D_p$ (\%) 
  & $u^2 \tau / D_p$ (\%)
 & $F$ (fN)
 & $\sqrt{\mathbf{S}}$ (fN)
 & $\Sigma$ (\%) 
  & $J$ ($k_B T$/s)\\
\hline
1 $\mu$m bare & 12.2 & 4 & 6.1 & 96 & 7 & 6.1 $\times 10^3$  \\
1 $\mu$m dec  & 29.4 & 33 & 9.5 & 154 & 16 & 9.3 $\times 10^2$ \\
200 nm bare & -7.4 & -- & -- & -- & -9 & --  \\
200 nm dec & -24 & -- & -- & -- & -50 & --  \\

\hline
\end{tabular}
\end{adjustbox}
\caption{\textbf{Enhancement via Optical Tweezer measurements.} Tabulated enhancement metrics for all conditions. Negative values indicate the metric was smaller for the active case compared to the passive case, likely due to aggregation. Entries marked (--) indicate a value that is not accessible.}
\label{tab:OT_enhancement_table}
\end{table}

\subsection{1 $\mu$m bare particles}

\begin{figure}[t]
    \centering
        \includegraphics[width=\linewidth]{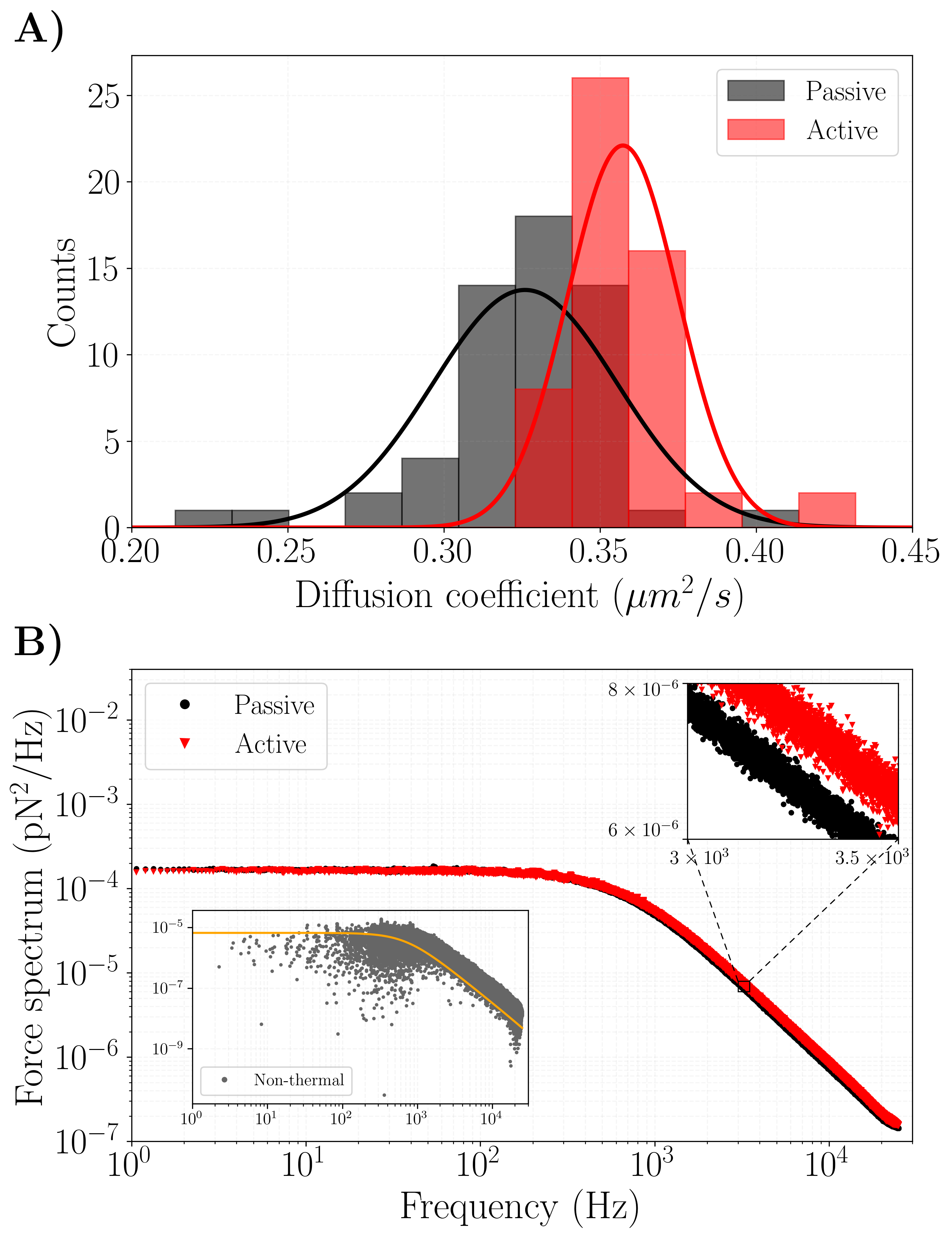}  
        \caption{\textbf{1 micron bare particles.} (A) Distribution of diffusion coefficients calculated from DDM for 1 $\mu$m bare particles in passive (0 U/mL urease) and active (50 U/mL urease) baths. The mean diffusion ($\pm$ standard deviation) in the passive bath ($n=56$) is $0.33 \pm 0.03~\mu$m$^2$/s, compared to $0.36 \pm 0.02~\mu$m$^2$/s in the active bath ($n=54$). (B) Force spectra of 1 micron bare particles in the active ($n=141$) and passive ($n=258$) case. Lower-left inset shows the nonthermal force spectrum, along with analytic model fit.  Upper-right zoomed inset shows the enhanced force fluctuations at high frequencies.}
        \label{fig:1umbare}
    \end{figure}

For 1 $\mu$m bare tracers, DDM measurements reveal a modest but reproducible diffusion enhancement in the presence of enzymatic activity. The mean diffusion increases from $0.33 \pm 0.01~\mu$m$^2$/s in the passive bath to $0.36 \pm 0.02~\mu$m$^2$/s in the active bath (Fig.~\ref{fig:1umbare}A), corresponding to an enhancement of $\Delta D / D_p \approx 10$\%. This level of enhancement corresponds to an effective driving force of $F \sim 5$ fN.

Optical tweezer (OT) measurements also show signatures of activity. The force spectra for active and passive conditions largely overlap at low frequencies, but deviations emerge at higher frequencies (Fig.~\ref{fig:1umbare}B), as highlighted in the inset. This indicates that activity-induced fluctuations are primarily concentrated in the high-frequency regime.

Quantitative estimates from OT depend on the metric used. Fitting the force spectra yields an enhancement in the range $u^2 \tau / D_p \sim 4$\% to $\Delta D / D_p \sim 12$\%, corresponding to an effective force of $F \sim 6$ fN. A model-free estimate based on integrating the nonthermal force spectrum gives $\sqrt{\mathbf{S}} \sim 96$ fN. The corresponding dissipation rate is $J \sim 6 \times 10^3~k_B T/\mathrm{s}$, and the variance-based force enhancement is $\Sigma \sim 7$\%.  Force distributions for passive and active are shown in Fig.~\ref{fig:hist_1µm}, left.

Taken together, these results indicate that enzymatic activity produces a modest but measurable effect for 1 $\mu$m bare tracers. The enhancement is detected by DDM and OT, with force spectra suggesting that the dominant contribution arises from short-timescale (high-frequency) fluctuations. This highlights both the limited persistence of activity in the bulk enzyme bath and the dependence of detectability on the measurement modality.

\subsection{1 $\mu$m decorated particles}

\begin{figure}[t]
    \centering
        \includegraphics[width=\linewidth]{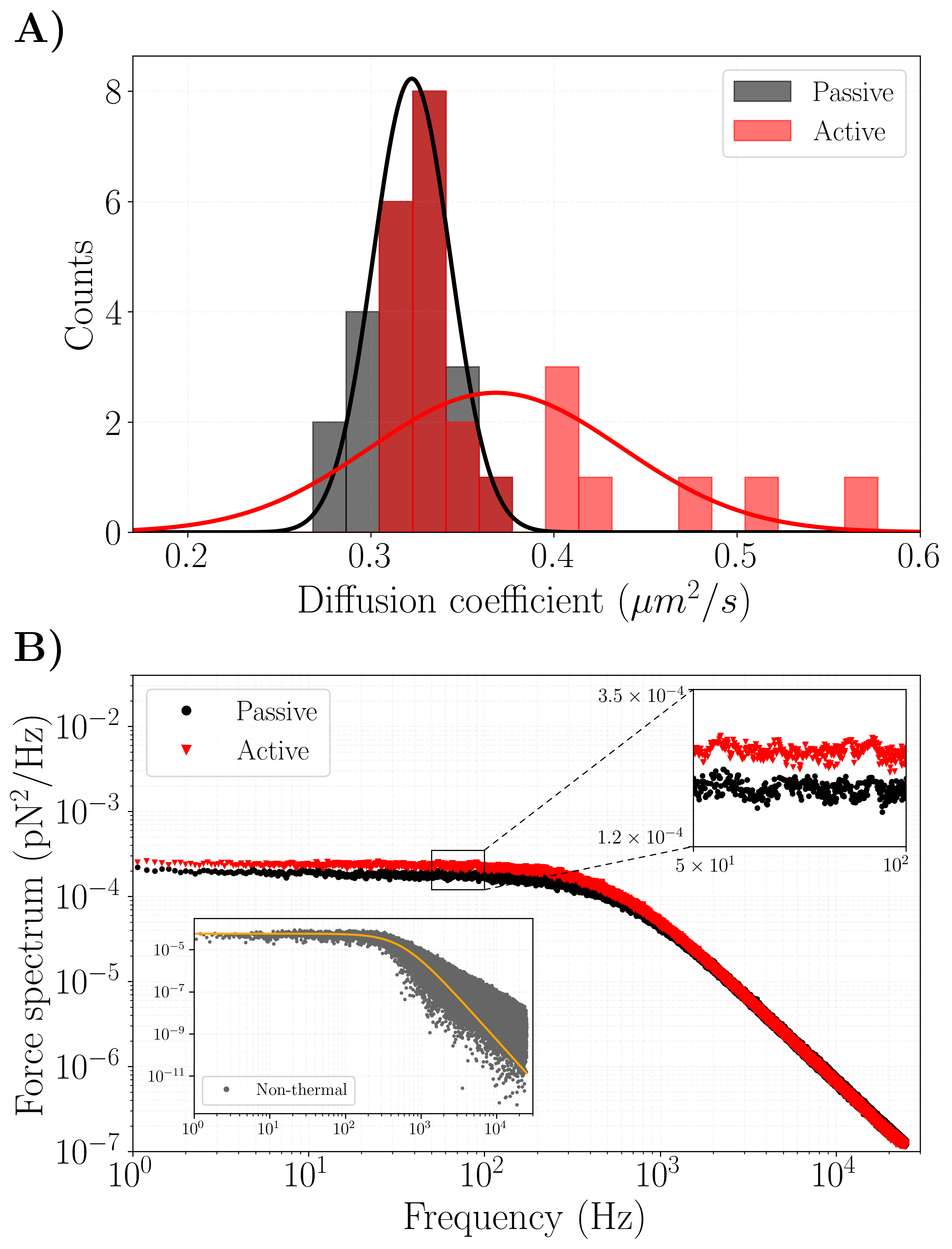}
    \caption{\textbf{1 micron decorated particles.} (A) Distribution of diffusion coefficients for 1 $\mu$m decorated particles in passive (0 U/mL urease) and active (50 U/mL urease) conditions.The mean diffusion ($\pm$ standard deviation) in the passive bath ($n=24$) is $0.32 \pm 0.022~\mu$m$^2$/s, compared to $0.37 \pm 0.071~\mu$m$^2$/s in the active bath ($n=24$). (B) Force spectra of 1 $\mu$m decorated particles in the active ($n=52$) and passive case ($n=29$). Lower-left inset shows the nonthermal force spectrum and analytic model fit.  Upper-right zoomed inset shows the enhanced force fluctuations that emerge primarily at lower frequencies.}
    \label{fig:1umdec}
\end{figure}

For 1 $\mu$m tracers directly decorated with urease we observed the strongest enhanced activity in this study. In DDM measurements, the mean diffusion increases from $0.32 \pm 0.022~\mu$m$^2$/s in the passive condition to $0.37 \pm 0.071~\mu$m$^2$/s in the active condition (Fig.~\ref{fig:1umdec}A), corresponding to an enhancement of $\Delta D / D_p \approx 14\%$. This corresponds to an effective driving force of $F \sim$ 7 fN.

Optical tweezer (OT) measurements show clear and robust signatures of activity. In contrast to bare tracers, the force spectra exhibit pronounced deviations at low frequencies (Fig.~\ref{fig:1umdec}B), indicating enhanced fluctuations on longer timescales. These low-frequency enhancements are well above experimental variability and are consistent with persistent, non-thermal forces acting on the particle. Notably, the dominant contribution shifts toward lower frequencies, closer to the characteristic timescale of the colloid rather than that of individual enzymes.

\begin{figure}[t]
    \centering
        \includegraphics[width=\linewidth]{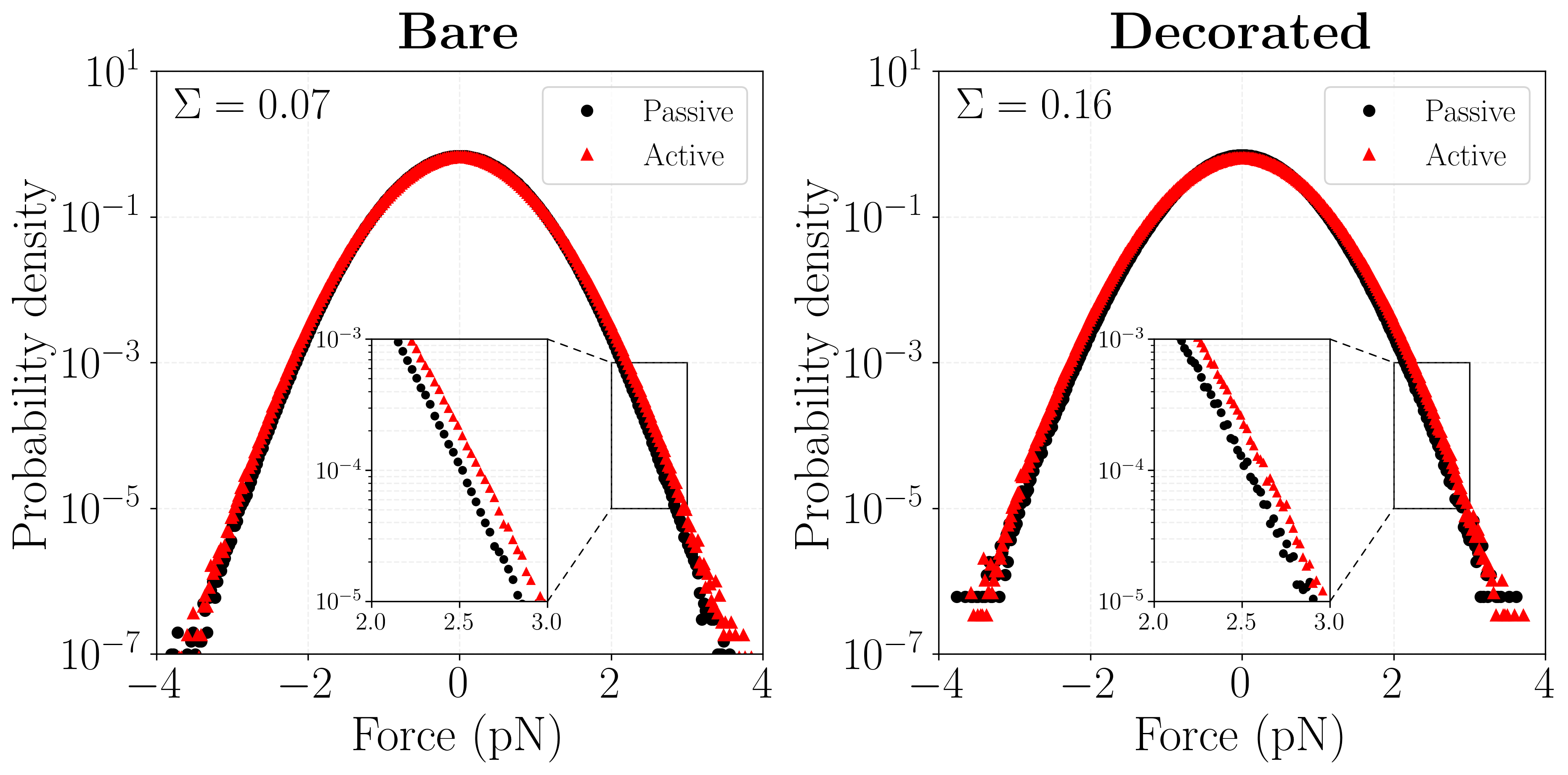}
    \caption{\textbf{Force distributions.} (Left) Force distributions for 1 $\mu$m bare particles in the active ($n=141$) and passive ($n=258$) case. (Right) Force distributions for 1 $\mu$m decorated particles in the active ($n=52$) and passive ($n=29$) case.  This variance-based metric for enhancement gives $\Sigma =$ 7\% and 16\% for active bath and active particle systems, respectively.}
    \label{fig:hist_1µm}
\end{figure}

Quantitative estimates from OT again depend on the metric used. Fitting the force spectrum yields an enhancement in the range $\Delta D / D_p \sim$ 29\% to $u^2 \tau / D_p \sim$ 33\%, corresponding to an effective force of $F \sim$ 10 fN. A model-free estimate based on integrating the nonthermal force spectrum gives $\sqrt{\mathbf{S}} \sim$ 154 fN. The corresponding dissipation rate is $J \sim 10^3$ $k_B T/\mathrm{s}$, and the variance-based enhancement is $\Sigma \sim$ 16\%. Force distributions for passive and active are shown in Fig.~\ref{fig:hist_1µm}, right.

Taken together, these results show that enzymatic activity localized at the tracer surface produces strong and persistent nonequilibrium fluctuations for 1 $\mu$m particles. In contrast to the bare case, where activity is weak and primarily high-frequency, the decorated particles exhibit substantial low-frequency enhancement that is more visible by both DDM and OT. This configuration therefore provides the clearest evidence that enzymatic activity can be effectively transduced into enhanced diffusion and force fluctuations at the micron scale.

\subsection{200 nm bare particles}

\begin{figure}[t]
    \centering
        \includegraphics[width=\linewidth]{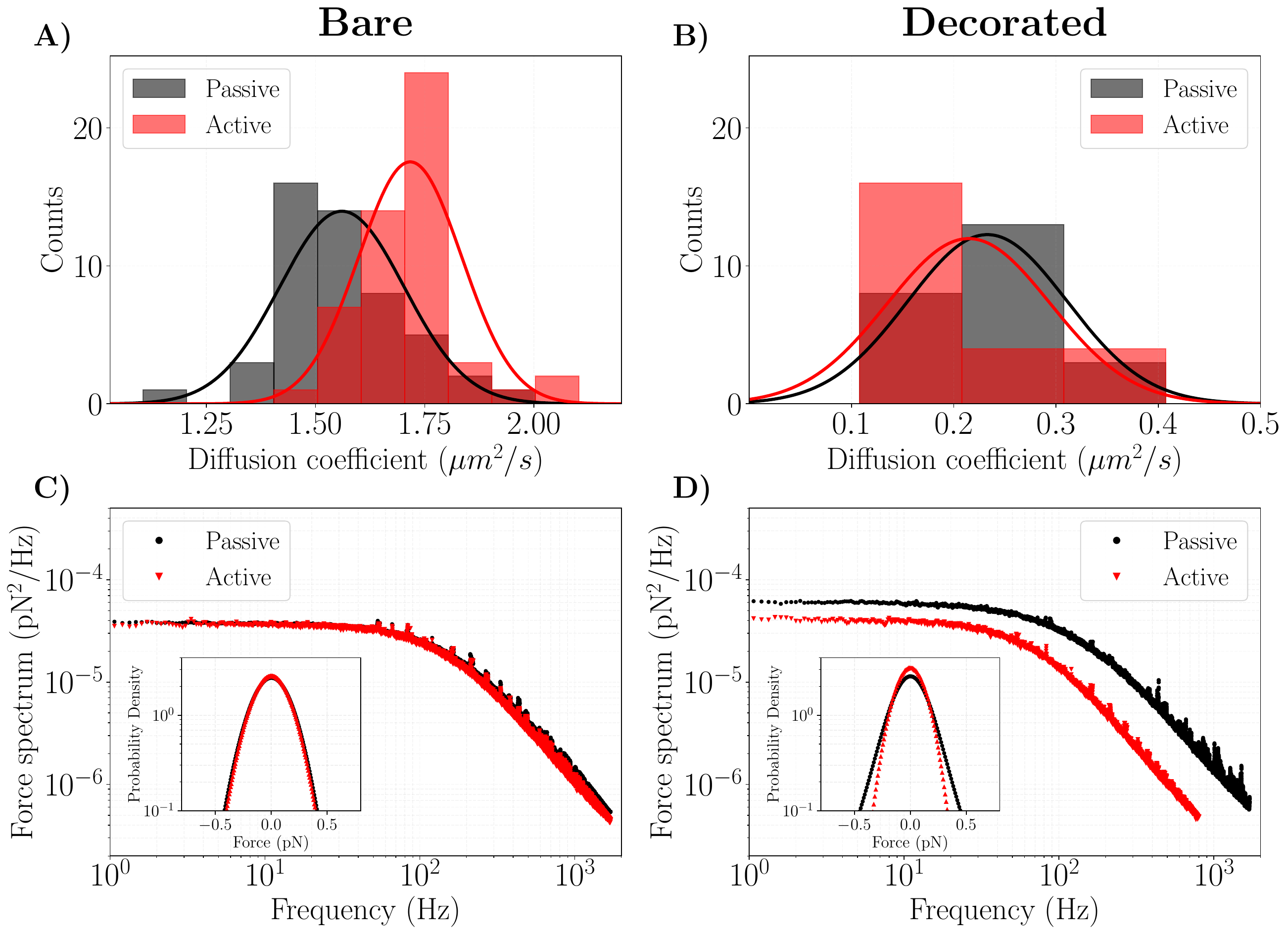}
    \caption{\textbf{200 nm bare and decorated particles}. (A, B)  Distribution of diffusion coefficients for 200 nm bare and decorated particles in passive (0 U/mL urease) and active (50 U/mL urease) conditions. The mean diffusion ($\pm$ standard deviation) of a bare particle in the passive bath ($n=50$)  is $1.56 \pm 0.14~\mu$m$^2$/s, compared to $1.72 \pm 0.12~\mu$m$^2$/s in the active bath ($n=52$). For the decorated particle in the passive condition (0 U/mL urease) the mean diffusion is 0.23$\pm 0.08~\mu^2$m/s ($n=24$) and for the active condition (50 U/mL urease) the mean value is 0.22$\pm 0.08~\mu^2$m/s ($n=24$).
    (C, D) Force spectra of 200 nm bare (active, $n = 82$; passive $n=494$) and decorated particles (active, $n=83$ ; passive $n=294$) do not exhibit enhancement. The inset in each spectrum shows the force distribution with $\Sigma = -9$ and $\Sigma = -50$ for bare and decorated, respectively.}
    \label{fig:200bare}
\end{figure}

For 200 nm bare tracers, DDM measurements reveal a modest diffusion enhancement in the presence of enzymatic activity. The mean diffusion increases from $1.57 \pm 0.17~\mu$m$^2$/s in the passive bath to $1.72 \pm 0.12~\mu$m$^2$/s in the active bath (Fig.~\ref{fig:200bare}A), corresponding to an enhancement of $\Delta D / D_p \approx$ 10\%. This corresponds to an effective driving force of $F \sim$ 28 fN.

In contrast, optical tweezer (OT) measurements do not show a clear signature of activity. The force spectra for active and passive conditions largely overlap across the measured frequency range (Fig.~\ref{fig:200bare}C), indicating that any activity-induced fluctuations are below the sensitivity of the measurement. Quantitative estimates from OT are therefore limited, since no enhancement can be detected. The small shift downward in the force spectra at high frequencies may reflect changes in the effective hydrodynamic radius. Even for nominally bare 200 nm particles, small aggregates are difficult to resolve optically, and seemed to occur more frequently in active baths. Any aggregation would increase the effective radius $R$. Since $\gamma = 6\pi\eta R$ and $\mu = \kappa/\gamma \propto 1/R$, an increase in $R$ would shift the trap response toward lower frequencies and reduce high-frequency spectral power. Thus, subtle spectral differences in the 200 nm measurements should be interpreted cautiously, as they may reflect unresolved changes in particle size.
Taken together, these results indicate that 200 nm bare tracers exhibit a modest enhancement in ensemble-averaged diffusion via DDM, but that this enhancement is not resolved in the OT force spectra. This does not necessarily imply that activity is absent. Rather, the activity-induced contribution may be too weak, too broadly distributed in frequency, or too small relative to the thermal and trap-mediated background to be resolved in single-particle force measurements.

\subsection{200 nm decorated particles}

\begin{figure}[th]
    \includegraphics[width = 0.45 \textwidth]{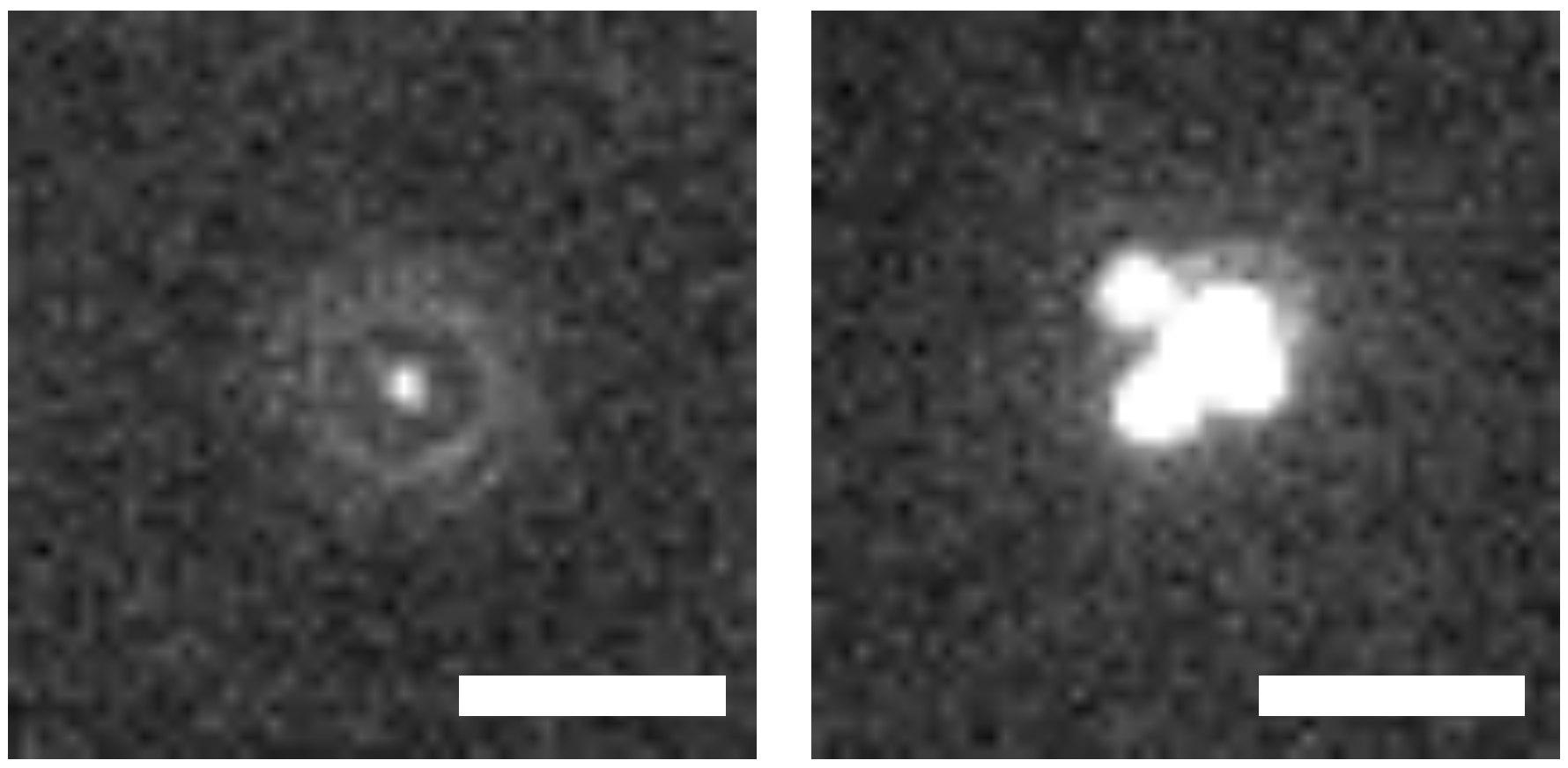}
    \caption{200 nm decorated particles exhibit clumping. Left is single particle, and right shows an aggregated clump of several particles. Scale bar = 2 $\mu$m}
    \label{fig:200nmclump}
\end{figure}

For 200 nm tracers directly decorated with urease, DDM measurements show no measurable diffusion enhancement. The mean diffusion coefficient is $0.23 \pm 0.08~\mu\mathrm{m}^2/\mathrm{s}$ in the passive condition and $0.22 \pm 0.08~\mu\mathrm{m}^2/\mathrm{s}$ in the active condition (Fig.~\ref{fig:200bare}B). Notably, the measured diffusion coefficients are substantially lower than both theoretical expectations and values obtained for bare 200 nm particles, which is puzzling at first. Microscopy reveals that these enzyme-decorated 200 nm tracers frequently aggregate into multi-particle clusters (Fig.~\ref{fig:200nmclump}). As a result, the measured dynamics  reflect the motion of these aggregates rather than individual particles, which may obscure potential activity-induced enhancement.

Optical tweezer (OT) measurements are consistent with the DDM results and show no evidence of enhanced force fluctuations. In fact, the force spectrum of the active condition is lower than that of the passive condition (Fig.~\ref{fig:200bare}D). While a similar effect is weakly visible for the 200 nm bare particles, the magnitude of the shift is substantially larger here, where microscopy clearly reveals the formation of multi-particle aggregates (Fig.~\ref{fig:200nmclump}). This suggests that the measured force spectra are dominated by aggregation-induced changes in the effective tracer properties rather than enzyme-induced fluctuations.

Taken together, neither DDM nor OT provides evidence for activity-induced enhancement in the 200 nm decorated system. Instead, the measurements appear to be dominated by aggregation, which is directly observed by microscopy and likely obscures any underlying enzyme-driven fluctuations. As a result, in our hands, this system does not provide a reliable platform for isolating activity-induced motion. In contrast to the 1~$\mu$m decorated particles, which remain well dispersed and exhibit clear signatures of activity, the 200 nm decorated particles highlight the importance of colloidal stability when probing enzyme-driven transport at the nanoscale.

\section{Discussion}

\subsection{Comparison of DDM and OT measurements}

DDM and OT provide complementary views of enzyme-enhanced tracer motion because they probe different observables over different timescales. DDM measures the displacements of freely diffusing particles and extracts an effective diffusion coefficient from the relaxation of density fluctuations. In our experiments, DDM is primarily sensitive to dynamics occurring over $10^{-1}$--$10^{2}$ Hz, set by the imaging frame rate (100 fps) and recording duration (20--30 s). OT, in contrast, directly measures force fluctuations acting on a trapped particle. For the trap stiffnesses and particle sizes used here, the characteristic trap response occurs at frequencies of order $10^2$--$10^3$ Hz, and force fluctuations are analyzed over an approximate frequency range of $10^{0}$--$10^{4}$ Hz. Thus, while the two techniques probe overlapping aspects of the dynamics, they emphasize different temporal windows and different physical observables. An additional distinction is that DDM measures freely diffusing particles, whereas OT confines the particle within a harmonic potential, altering its response to external fluctuations.

Despite these differences, the two methods show broadly similar trends. Both DDM and OT detect activity-induced enhancement for the 1~$\mu$m bare particles and, more strongly, for the 1~$\mu$m decorated particles. The largest enhancement is observed for the decorated micron-scale particles, where both increased diffusion and pronounced excess force fluctuations are detected. The primary discrepancy occurs for the 200 nm bare particles, where DDM reveals a diffusion enhancement of approximately 10\%, while OT does not show a corresponding increase in force fluctuations. These observations are not necessarily contradictory. DDM and OT quantify different physical observables over different temporal windows, so identical values are not expected even when both probe the same underlying activity. Fluctuations that contribute to long-time transport may produce only weak changes in the measured force spectrum, particularly when the particle is confined by the optical trap. 

A similar distinction has been reported previously for enzyme-powered colloids. Ma \emph{et al.} measured an effective propulsion force of $\sim64$~fN for catalase-powered Janus nanoparticles using optical tweezers, whereas the corresponding force estimated from long-time diffusion was approximately a factor of two smaller ($\sim 32$~fN)~\cite{ma2015enzyme}. They attributed this difference to OT probing instantaneous single-particle force fluctuations while diffusion-based methods measure a time-averaged effective response. Our observations are consistent with this interpretation. Rather than representing conflicting measurements, DDM and OT provide complementary information about different aspects of the underlying nonequilibrium dynamics. In the present study, both methods identify the same qualitative trend: the strongest activity-induced enhancement occurs for 1~$\mu$m urease-decorated particles, weaker enhancement for 1~$\mu$m bare particles.

\subsection{Comparison with enzyme-powered colloid force measurements}

The force scales measured here are comparable to previous optical tweezer measurements on enzyme-powered colloids, despite substantial differences in enzyme type, particle architecture, and analysis methods. Ma \emph{et al.} reported an effective force of $64$~fN for catalase-powered Janus hollow mesoporous silica nanoparticles ($\sim$390 nm) using a power spectral density analysis~\cite{ma2015enzyme}. Patiño \emph{et al.} measured $170$~fN for 2~$\mu$m urease-powered PS@SiO$_2$ micromotors using a fluctuation-based analysis of trapped particle displacements~\cite{patino2018influence}. In comparison, our model-independent integration of the nonthermal force spectrum yields $\sqrt{\mathbf{S}}\approx96$~fN and $154$~fN for 1~$\mu$m bare and decorated polystyrene particles, respectively, placing our measurements within the same quantitative range.

But, direct quantitative comparison should be made with caution because the systems differ in enzyme identity, fuel, particle composition, enzyme loading, and force metric. Nevertheless, an important qualitative comparison emerges. Patiño \emph{et al.} observed a threshold in enzyme coverage below which no measurable propulsion force was detected and above which forces increased to $\sim170$ fN~\cite{patino2018influence}. They also reported measurable propulsion only for urease-functionalized PS@SiO$_2$ particles, whereas comparable polystyrene particles showed little or no enhancement. Our urease-decorated particles are based on a polystyrene substrate rather than a rough silica architecture, making them more directly comparable to their polystyrene control particles than to the PS@SiO$_2$ micromotors. Despite this, we observe measurable activity-induced force fluctuations. While the magnitude is somewhat lower than that reported for PS@SiO$_2$, our results suggest that detectable enzyme-driven fluctuations do not require rough silica architectures, although increased surface area and enzyme loading likely enhance the response.

\begin{figure}[t]
    \centering
    \includegraphics[width=0.45\textwidth]{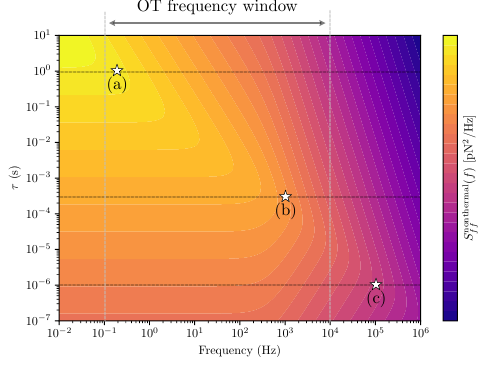}
    \caption{
    \textbf{Persistence time determines the observability of active force fluctuations.}
     Predicted nonthermal force spectrum from the AOUP model as a function of persistence time, $\tau$, and frequency for a fixed active speed. The gray vertical dashed lines indicate the approximate frequency range accessible in our OT measurements. White stars denote the characteristic crossover frequency of the nonthermal force spectra, $f_c=(2\pi\tau)^{-1}$, for three representative persistence times: (a) $\tau \sim 1$ s, representative of micron-scale active colloids; (b) $\tau \sim 3\times10^{-4}$ s, comparable to the characteristic persistence extracted from our AOUP fits for 1~$\mu$m decorated particles; and (c) $\tau \sim 10^{-6}$ s, representative of freely diffusing enzymes in solution. As persistence decreases, the crossover shifts to progressively higher frequencies, shifting features of the nonthermal force spectrum outside the experimentally accessible bandwidth and making them increasingly difficult to distinguish from the thermal background. Color bar indicates amplitude of nonthermal spectra increasing from purple to yellow.
    }
    \label{fig:AOUP_tau}
\end{figure}

\subsection{Persistence time and detectability of activity}

A key difference between active baths and active particles is the persistence time of the active forcing. For freely diffusing enzymes, the orientation of any activity-induced force rapidly decorrelates through rotational diffusion — for a nanometer-scale protein, the rotational timescale is on the order of $10^{-6}$~s. When enzymes are instead immobilized on a colloidal tracer, force generation becomes coupled to the much slower rotational dynamics of the particle, increasing the persistence time by several orders of magnitude. Importantly, localization does not increase the force generated per catalytic event; it increases the time over which that force remains directionally correlated.

This distinction has direct consequences for what is experimentally observable. Within the AOUP framework, persistence time controls the shape of the nonthermal force spectrum through the second term of equation~(\ref{eq:forcespectrum}): the characteristic crossover frequency $f_c=(2\pi \tau)^{-1}$ marks the transition between a low-frequency plateau of nearly constant active fluctuations and a high-frequency regime where the spectrum decays rapidly. As $\tau$ decreases, this crossover shifts to higher frequencies, progressively pushing active fluctuations outside the experimentally accessible bandwidth.

Fig.~\ref{fig:AOUP_tau} illustrates this effect across a broad range of persistence times, with three representative cases highlighted. For long persistence ($\tau \sim 1$~s; Fig.~\ref{fig:AOUP_tau}(a)), typical of slowly rotating micron-scale colloids, the crossover lies near the lower edge of our OT measurement window, producing a strong frequency-dependent signature in the force spectrum. At intermediate persistence ($\tau \sim 3\times10^{-4}$~s; Fig.~\ref{fig:AOUP_tau}(b)), consistent with values extracted from our AOUP fits to the 1~$\mu$m decorated particles, the crossover falls squarely within the accessible frequency range, yielding a pleateau at lower frequencies but strong variation in the higher frequency regime of the accessible window. For freely diffusing enzymes ($\tau \sim 10^{-6}$~s; Fig.~\ref{fig:AOUP_tau}(c)), however, the crossover occurs near $10^5$~Hz — well above our measurement bandwidth — so only the flat, low-frequency tail of the active spectrum is sampled, making activity-induced fluctuations difficult to distinguish from thermal noise.

This framework offers a straightforward interpretation of the experimental trends. The 1~$\mu$m decorated particles show the clearest excess force fluctuations because immobilizing enzymes on their surface increases the persistence of the active forcing, shifting spectral power into the accessible frequency range. The weaker signatures for 1 $\mu m$ bare particles are consistent with rapidly decorrelating, freely diffusing enzymes. For 200~nm particles, two compounding effects suppress detectability: smaller particles rotate approximately 125$\times$ faster than 1~$\mu$m particles (since $D_r \propto R^{-3}$), further shifting the active spectrum to high frequencies; and aggregation of the 200~nm decorated particles additionally complicates force measurements by altering the effective hydrodynamic radius and trap response.

More broadly, these results establish persistence time as a key parameter governing the observability of enzyme-driven fluctuations. Systems with comparable active forcing but different persistence times can produce qualitatively different force spectra and diffusion signatures. Localizing enzymatic activity to a colloidal tracer therefore provides a simple physical mechanism for bringing active fluctuations into experimentally accessible timescales — enhancing their detectability without necessarily increasing the underlying energy input.

\subsection{Implications for enzyme-driven transport}

Our results connect two previously somewhat separate bodies of work: studies of 
enhanced diffusion in active enzyme solutions, and the broader active matter 
literature on force transduction from microscopic activity to larger tracers. 
Previous work by Zhao~\emph{et al.}~\cite{zhao2017enhanced} demonstrated that 
passive tracers can exhibit $\sim$20\% diffusion enhancement when suspended in 
active enzyme solutions, suggesting that enzymatic activity influences the dynamics 
of surrounding particles even when those particles are chemically inert. Our 
observations for bare particles are consistent with this picture. More generally, 
recent reviews have emphasized that such effects are best understood within an 
active matter framework, rather than solely as a question of enzyme 
self-propulsion~\cite{zhang2019enhanced,tripathi2022gauging,astumian2026biomolecular}.

A finding of this work is that localizing enzymatic activity directly to 
the tracer surface produces a qualitatively different dynamical response compared 
to dispersing passive tracers in an active enzyme bath. This is consistent with 
the micromotor literature, where particle architecture, enzyme identity, and enzyme 
loading are known to strongly influence transport and force 
generation~\cite{ma2015enzyme,patino2018influence,patino2019self,paffen2025programmable}. 
Our results extend this picture to comparatively simple urease-decorated polystyrene 
particles, demonstrating that measurable enzyme-induced force fluctuations do not 
require complex particle fabrication.

The physical origin of this distinction lies in the persistence of active 
fluctuations, as discussed in the previous section. Active Brownian particle and 
AOUP theories predict that transport enhancements depend on both the strength 
and persistence time of the active 
forcing~\cite{fodor2018statistical,martin2021statistical,tripathi2022gauging}. 
Bare particles in active enzyme baths experience rapidly decorrelating fluctuations 
that manifest primarily as modest increases in high frequency fluctuations that accumulate into long-time diffusion. When activity 
is localized to the tracer surface, fluctuations become more persistent, shifting 
spectral power to lower frequencies and producing signatures detectable by both 
DDM and OT. The motion of larger tracer particles can thus serve as a probe of the 
underlying active fluctuations and their persistence, an idea that has also been 
explored theoretically in the context of nanoswimmers~\cite{tripathi2022gauging}.

Taken together, these results underscore that activity-induced enhancement is not 
a single universal quantity, but depends on how microscopic fluctuations are 
transduced across length scales and on the temporal window over which they are 
measured. This perspective may help reconcile the variability in reported 
enzyme-enhanced diffusion across experimental 
systems~\cite{zhang2019enhanced,feng2020enhanced,presse2020thermodynamic}: 
differences between studies likely reflect not only variations in enzyme systems 
and sample preparation, but also differences in persistence time, tracer size, 
activity localization, and measurement technique. A complete picture of 
enzyme-driven transport therefore requires considering force fluctuations, 
long-time diffusion, and persistence together within a unified dynamical 
framework.

\section{Conclusion}

Using complementary measurements of long-time diffusion and short-time force fluctuations, we show that enzymatic activity can generate measurable enhancements in the dynamics of colloidal tracers, with the strongest effects observed when activity is localized directly at the particle surface. While bare particles in active enzyme baths exhibit only modest enhancements, enzyme-decorated micron-scale particles display clear signatures of nonequilibrium activity in both DDM and OT measurements. Together, our results suggest that the transduction of enzymatic activity to larger length scales depends not only on the magnitude of the underlying fluctuations, but also on their persistence and how they are probed experimentally, highlighting the importance of considering timescale and observable when interpreting enzyme-driven transport.

\section*{Acknowledgements}
This material is based upon work supported by the National Science Foundation under Grant No. NSF DMR-2004566 to WWA, NSF DMR-2004417 to JLR, and NSF DMR- 2004400 to WBR.  WWA also acknowledges funding from ANR Grant No. ANR-23-CPJ1-0170-01 and the Erskine Fellowship Program at the University of Canterbury. MG and EL were partially supported by the Dan Black Family Trust Fellowship. MG was partially supported by the Eiker-Adams Fellowship. JMP was supported by a doctoral fellowship from the French Ministry of Higher Education, Research and Innovation (MESRI).

\pagebreak{}


\bibliographystyle{unsrt}
\bibliography{bibtex}

\end{document}